\documentstyle[epsf,aps,psfig,12pt]{revtex}
 
\begin{document} 

\begin{center}
{\Large \bf Massive quark propagator and competition between 
chiral and diquark condensate } \\[3mm]
Mei Huang$^{1}$, Pengfei Zhuang$^{1}$, Weiqin Chao$^{2}$ \\[2mm]
$^1$ Physics Department, Tsinghua University, Beijing 100084, 
China \\
huangmei@mail.tsinghua.edu.cn \\
zhuangpf@mail.tsinghua.edu.cn \\[2mm]
$^2$ CCAST, Beijing 100080, China \\ 
Institute of High Energy Physics, CAS, Beijing 100039, China \\
chaowq@hp.ccast.ac.cn
\end{center}

\date{}

\begin{abstract}

The Green-function approach has been extended to the moderate 
baryon density
region in the framework of an extended Nambu--Jona-Lasinio model, 
and the thermodynamic potential with both chiral and diquark 
condensates 
has been evaluated by using the massive quark propagator. The 
phase structure 
along the chemical potential direction has been investigated 
and the strong 
competition between the chiral and diquark condensate has been 
analyzed by 
investigating the influence of the diquark condensate on the 
sharp Fermi surface. 
The influence of the diquark condensate on the quark properties 
has been 
investigated, even though the quarks in the color breaking phase are very 
different from that in the chiral breaking phase, the difference between 
quarks in different colors is very small.   

\end{abstract}

\newpage

\section{Introduction}

QCD phase transitions along the baryon density direction has
attracted much attention recently since it was found that the 
color-superconducting gap can be of the order of $100 {\rm MeV}$
\cite{raja1} and \cite{Rapp1}, which is two orders larger than 
early perturbative estimations \cite{love}.  However, till now, there is 
no uniform framework discussing the phase structure in the wide 
region of chemical potential $\mu$ from about $300 MeV$ to $10^8 MeV$. 

In the idealized case at asymptotically high baryon densities, 
the color superconductivity with two massless flavors and the 
color-flavor-locking 
(CFL) phase with three degenerate massless quarks have been widely 
discussed from first principle QCD calculations, see \cite{raja2} and references 
therein. Usually, the diagrammatic methods are used in the 
asymptotic 
densities. The Green-function of the eight-component field 
and the gap equation 
were discussed in details in \cite{rischke1} \cite{rischke2}. 
Neither current
quark mass nor chiral condensate are necessary to be 
considered because they can be 
neglected compared with the very high Fermi surface.
At less-than-asymptotic densities, the corrections of 
non-zero quark mass 
to the pure CFL phase can be treated perturbatively by 
expanding the current quark mass around the chiral limit 
\cite{massschafer4} \cite{rho2}. 

For physical applications, we are more interested in the 
moderate baryon density region, which may be related to the neutron 
stars and, in very optimistic cases even to heavy-ion collisions.
Usually, effective models such as instanton, as well as 
Nambu-Jona-Lasinio (NJL) model, are used. The model parameters are fixed 
in the QCD vacuum. In this region, the usual way is to use the variational methods 
working out the gap equations from the thermodynamic potential 
\cite{raja3}-\cite{ebert3}, except for \cite{carter} and 
\cite{rapp2} in the instanton model, where the quark propagator was 
evaluated explicitly, but the form is complicated.

One of our main aims in this paper is to apply the Green-function approach 
in the moderate baryon density region. To work out the phase structure from 
hadron phase to the color superconducting phase, one should deal with the chiral 
condensate and diquark condensate simultaneously. Because the chiral 
condensation contributes a dynamic quark mass, it is not reasonable any more in this 
density region to treat the quark mass term perturbatively, like in \cite{rho2}. 
By using the energy projectors for massive quark, we will evaluate the Nambu-Gorkov
massive quark propagator, which will help us deal with the chiral and 
diquark condensate simultaneously.

In the normal phase, the quarks in different colors are degenerate, while in the 
color breaking phase, it is natural to assume that the quarks 
involved in the diquark condensate are
different from that not participating in the diquark condensate.  
In \cite{carter} and \cite{rapp2}, different masses for the 
quarks which participate 
and not participate in the diquark condensate are introduced. However, we will 
see it difficult to get the mass expression for the quarks participating in the 
diquark 
condensate, because the particles and holes mix with each other and the elementary 
excitations are quasi-particles and quasi-holes near the Fermi surface. 
In our case, the difference between quarks in different colors has been reflected by 
their propagators, we read the difference through calculating the 
quarks' chiral condensate, but not trying to work out their masses. 
In the chiral limit, the chiral condensate disappears entirely in the color 
superconducting phase, it is not possible to investigate the influence of 
color breaking on quarks in different colors, so we will keep the current quark 
mass finite in this paper.
   
In the moderate baryon density region, people are interested in 
the question whether there exists a region where both the chiral 
symmetry and color symmetry are broken \cite{raja3}-\cite{kerbikov}.  
In the chiral limit it was found that the existence of the mixed broken phase depends 
on the coupling constants $G_S$ and $G_D$ in the quark anti-quark 
and diquark channels \cite{klevansky}\cite{vander}.  
In the case of small ratio of $G_D/G_S<1$ the calculations in the
instanton model \cite{raja3}\cite{berges2}, NJL model \cite{klevansky}
and random matrix model \cite{vander} show a strong competition 
between the chiral and diquark condensates, i.e., 
where one condensate is nonzero the other vanishes.
While for $G_D/G_S>1$ the calculations in the random matrix model 
\cite{vander} and NJL model \cite{klevansky} show that there exists
a region where both chiral (dynamical) symmetry and color symmetry 
are broken, and the chiral and diquark condensate coexist.
The larger value of $G_D/G_S$ is, the wider region of the mixed broken
phase has been found in the random matrix model \cite{vander}. 
 
The presence of a small current quark mass will induce that chiral symmetry 
only restores partially and there will always exist a mall chiral condensate 
in the color superconducting phase, this phenomena had been called 
coexistence of chiral and diquark condensates in \cite{raja3}\cite{berges2}. 
In order to differ from this coexistence of the diquark and the small chiral condensate
induced by current quark mass, we will call the coexistence region of the diquark
and large chiral condensate induced by large $G_D/G_S$ before the
chiral phase transition as mixed broken phase. 

In the coexistence region resulted by current quark mass, the chiral condensate
is small comparing with the diquark condensate, and the role of the chiral 
condensate can be prescribed by the Anderson theroem \cite{kerbikov}, i.e.,
in this phase, the contribution of the chiral condensate to 
thermodynamic quantities becomes strongly suppressed, and one can 
calculate the diquark condensate neglecting the influence of 
the chiral condensate.
  
In this paper, we will explain the existence of the mixed broken phase induced 
by larger $G_D/G_s$ by analyzing the influence of the diquark condensate 
on the Fermi surface. In the mean-field approximation of the NJL model, 
the thermal system of the constituent quarks is a nearly ideal Fermi gas, 
and there is a sharp Fermi surface. The chiral symmetry begins to 
restore when the chemical potential is larger than the quark mass 
in the vacuum. When a diquark condensate is formed, the Cooper pair extends 
the Fermi surface, which induces the chiral symmetry restoring 
at a smaller chemical potential.
The stronger the coupling constant in the diquark channel, 
the larger the diquark
condensate and the smoother the Fermi surface. 

In the following, we briefly introduce the extended NJL 
model in Sec. II, 
then in Sec III, we evaluate the thermodynamic potential 
using the massive 
quark propogator.  In Sec. IV, we get the gap equations and condensates. 
The numerical results and conclusions are given at the end.

\section{The extended NJL model}

The choice of the NJL model \cite{Nambu} is motivated by the fact that this 
model displays the same symmetries as QCD and that it describes 
well the spontaneous breakdown of chiral symmetry in the vacuum 
and its restoration at high temperature and density. 
The model we used in this paper is an extended version of the two-flavor NJL model 
including interactions in the color singlet quark-antiquark channel and 
color anti-triplet diquark channel, which is not directly extended from the 
NJL model, but from the QCD Lagrangian \cite{dnjl1}- \cite{dnjl3}.

The importance of color ${\bar 3}$ diquark degree of freedom is related to 
the fact that one can construct a color-singlet nucleon current based on it. Because 
the gluon exchange between two quarks in the color ${\bar 3}$ channel 
is attractive, one can view a color singlet baryon as a quark-diquark bound state.
And experimental data from $pp$ collisions indicate the existence of this 
quark-diquark component in nucleons.

The first attempt to investigate the diquark properties in the NJL model was taken in 
\cite{Vogl1}. Starting from an NJL model for scalar, pseudoscalar, vector and 
axial-vector interactions of the $({\bar q} q) \times ({\bar q} q)$ type  
and Fierz-transforming away the vector and axial-vector interactions, 
the scalar and pseudoscalar mesons, and diquarks can be obtained.  
However, this method could not get a consistent treatment of vector 
and axial-vector particles. 

The extended NJL model we used was derived directly from QCD Lagrangian
\cite{dnjl1} \cite{dnjl2}. Integrating out gluon degrees of freedom from
QCD Lagrangian,  and performing a local approximation for the (nonperturbative)
gluon propagator, one gets a contact current-current interaction.
By using a special Fierz-rearrangement \cite{cahill}, one can completely 
decompose the two-quark-current interaction term into "attractive" color 
singlet (${\bar q}q$) and color antitriplet ($qq$) channels. 
In this way, a complete simultaneous description of scalar, 
pseudo-scalar, vector, 
and axial-vector mesons and diquarks is possible, thus the extended NJL model 
including $({\bar q} q) \times ({\bar q} q)$ interactions is completed by 
a corresponding 
$({\bar q} {\bar q}) \times (q q)$ interaction part.

In our present work, we only consider scalar, pseudoscalar mesons and 
scalar diquark, and using the following Lagrangian density
\begin{eqnarray}
\label{lagr}
{\cal L} = {\bar q}(i\gamma^{\mu}\partial_{\mu}-m_0)q + 
   G_S[({\bar q}q)^2 + ({\bar q}i\gamma_5{\bf {\vec \tau}}q)^2 ]
 +G_D[(i {\bar q}^C  \varepsilon  \epsilon^{b} \gamma_5 q )
   (i {\bar q} \varepsilon \epsilon^{b} \gamma_5 q^C)],
\end{eqnarray}
where $q^C=C {\bar q}^T$, ${\bar q}^C=q^T C$ 
are charge-conjugate spinors, $C=i \gamma^2 \gamma^0$ is the charge 
conjugation matrix (the superscript $T$ denotes the transposition operation),
$m_0$ is the current quark mass, the quark field $q \equiv q_{i\alpha}$ 
with $i=1,2$ and $\alpha=1,2,3$ is a flavor 
doublet and color triplet, as well as a four-component Dirac spinor, 
${\vec \tau}=(\tau^1,\tau^2,\tau^3)$ are Pauli matrices in the flavor 
space, where $\tau^2$ is antisymmetric, and $(\varepsilon)^{ik} \equiv \varepsilon^{ik}$,
$(\epsilon^b)^{\alpha \beta} \equiv \epsilon^{\alpha \beta b}$ are totally 
antisymmetric tensors in the flavor and color spaces.  

In (\ref{lagr}), $G_S$ and $G_D$ are independent effective coupling constants in the 
scalar quark-antiquark and scalar diquark channel. The former is responsible 
for the meson excitations, and the later for the diquark excitations, which 
in principle can be determined by fitting mesons' and baryons' properties 
in the vacuum. The attractive interaction in different channels in this Lagrangian will 
give rise to a very rich structure of the phase diagram. At zero temperature and density, 
the attractive interaction in the color singlet channel is responsible for the 
appearance of a quark anti-quark condensate and for the spontaneous breakdown 
of the chiral symmetry, and the interaction in the $qq$ channel 
binds quarks into diquarks (and baryons), but is not strong enough
to induce diquark condensation. As the density increases, Pauli blocking 
suppresses the ${\bar q}q$ interaction, while the attractive interaction in the 
color anti-triplet diquark channel will induce the quark-quark condensate around 
the Fermi surface which can be identified as a superconducting phase.  

After bosonization \cite{dnjl1}\cite{dnjl2}, one can obtain the linearized version of the 
model (\ref{lagr}) 
\begin{eqnarray}
\label{lagr2}
\tilde{\cal L} & =  & {\bar q}(i\gamma^{\mu}\partial_{\mu}-m_0)q - 
  {\bar q}(\sigma+i \gamma^5{\vec \tau}{\vec \pi}) q - 
  \frac{1}{2}\Delta^{*b} (i{\bar q}^C  \varepsilon \epsilon^{b}\gamma_5 q )
  -\frac{1}{2}\Delta^b (i {\bar q}  \varepsilon  \epsilon^{b} \gamma_5 q^C) \nonumber \\
  & & -\frac{\sigma^2+{\vec \pi}^2}{4G_S}
  -\frac{\Delta^{*b}\Delta^{b}}{4G_D},
\end{eqnarray}
with the bosonic fields 
\begin{eqnarray}
\Delta^b \sim i {\bar q}^C \varepsilon \epsilon^{b}\gamma_5 q, \ \ 
\Delta^{*b} \sim i {\bar q}  \varepsilon  \epsilon^{b} \gamma_5 q^C, \ \
\sigma \sim {\bar q} q , \ \ {\vec \pi} \sim i {\bar q}\gamma^5 {\vec \tau} q.
\end{eqnarray}

Clearly, the $\sigma$ and ${\vec \pi}$ fields are color singlets, and the 
diquark fields $\Delta^b$ and $\Delta^{*b}$ are color antitriplet and  
(isoscalar) singlet under the chiral $SU(2)_L \times SU(2)_R$ group.
$\sigma \neq 0$  and $\Delta^b \neq 0$ indicate that chiral symmetry and color
symmetry are spontaneously broken. 
We assume that the two condensates coexist, i.e., 
\begin{eqnarray}
\label{meanfield}
 \sigma  & \neq & 0, ~ {\vec \pi}  = 0; \nonumber \\
 \Delta^1 & = &  \Delta^2  = 0,  \Delta^3 \neq 0,
\end{eqnarray}
Here it has been regarded that only the first two
colors participating in the condensate, while the third one does not. In the
later expressions, we will simply use $\Delta \equiv \Delta^3$.

The real vacuum will be determined by the minimum of the thermodynamic 
potential at $T=0$ and $\mu=0$, and the minimum of the thermodynamic 
potential at any $T, \mu$ determines the stable state at that point.

\section{Partition function and thermodynamic potential}

\subsection{Nambu-Gorkov formalism}

The partition function of the grand canonical ensemble can be evaluated 
by using standard method \cite{kapusta} \cite{bellac},
\begin{eqnarray}
\label{part}
{\cal Z}=N' \int [d {\bar q} ][d q] exp\{ \int_0^{\beta} d \tau \int d^3{\vec x} 
  ~ ( \tilde{\cal L} +\mu {\bar q} \gamma_0 q)\},
\end{eqnarray}
where $\mu$ is the chemical potential, and $\beta=1/T$ is the inverse of temperature $T$.

According to the mean field approximation Eq. (\ref{meanfield}), we can  
write the partition function as a product of three parts, 
\begin{eqnarray}
\label{part}
{\cal Z}={\cal Z}_{const} {\cal Z}_{q_{1,2}}{\cal Z}_{q_3} .
\end{eqnarray}
The constant part is
\begin{eqnarray}
{\cal Z}_{const}=N'{\rm exp} \{- \int_0^{\beta} d \tau \int d^3{\vec x} ~ 
[\frac{\sigma^2}{4 G_S}
      +\frac{\Delta^{*}\Delta}{4 G_D} \}.
\end{eqnarray}
For the quarks in the first two colors (named as "the first two quarks" in the following) 
$Q=q_{1,2}$ participating in the quark condensate, one has
\begin{eqnarray}
\label{z12}
{\cal Z}_{q_{1,2}} & = & \int[d{\bar Q}][d Q] {\rm exp} \{ \int_0^{\beta} d \tau \int d^3{\vec x} ~ 
[\frac{1}{2}{\bar Q} 
( i\gamma^{\mu}\partial_{\mu}-m+\mu)Q + \nonumber \\
 & & \frac{1}{2}{\bar Q}^C 
( i\gamma^{\mu}\partial_{\mu}-m-\mu)Q^C 
+\frac{1}{2}{\bar Q} \Delta^{-} Q^C + \frac{1}{2}{\bar Q}^C \Delta^{+} Q] \}.
\end{eqnarray}
Here we  have introduced the constituent quark mass 
\begin{eqnarray}
m=m_0+ \sigma,
\end{eqnarray}
and defined $\Delta^{\pm}$ as
\begin{eqnarray} 
\Delta^{-} = -i \Delta \varepsilon  \epsilon^{b} \gamma_5, \  \ 
\Delta^{+} = -i \Delta^{*} \varepsilon  \epsilon^{b} \gamma_5
\end{eqnarray}
with the relation $\Delta^{+}= \gamma^0 (\Delta^{-})^{\dagger} \gamma^0$.

For the quark in the third color (named as "the third quark" in the following),
which is not involved in the diquark condensate, one has 
\begin{eqnarray}
\label{z3}
{\cal Z}_{q_3}  =  \int[d{\bar q}_3][d q_3]{\rm exp}\{ \int_0^{\beta} d \tau 
\int d^3{\vec x} ~[\frac{1}{2}{\bar q}_3 ( i\gamma^{\mu}\partial_{\mu}-m+\mu)q_3
+ \frac{1}{2}{\bar q}_3^C ( i\gamma^{\mu}\partial_{\mu}-m-\mu)q_3^C] \}.
\end{eqnarray}

Introducing the 8-component spinors for the third quark and the first
two quarks, respectively
\begin{equation}
\Psi_3 =\left(\begin{array}{c} 
                    q_3 \\
                    q_3^C
                   \end{array}
            \right), \ \ 
\bar{\Psi}_3 =( \bar{q}_3 \ \  \bar{q}_3^C ),
\end{equation}
\begin{equation}
\Psi =\left(\begin{array}{c} 
                    Q \\
                    Q^C 
                   \end{array}
            \right), \ \ 
\bar{\Psi} =( \bar{Q} \ \  \bar{Q}^C ),
\end{equation}
and using the Fourier transformation in the momentum space,
\begin{eqnarray}
q(x)=\frac{1}{\sqrt{V}}\sum_n\sum_{\vec p} 
e^{-i ( \omega_n\tau-{\vec p}\cdot{\vec x})}q({\vec p}),
\end{eqnarray}
where $V$ is the volume of the thermal system,
we can re-write the partition function Eqs. (\ref{z12}) and (\ref{z3}) 
in the momentum space as
\begin{eqnarray}
\label{zq12}
{\cal Z}_{q_{1,2}} & = & \int[d \Psi]{\rm exp} \{\frac{1}{2}  \sum_{n,{\vec p}} ~
{\bar \Psi}\frac{{\rm G}^{-1}}{T}  \Psi \}  \nonumber \\
& = & {\rm Det}^{1/2}(\beta {\rm G}^{-1}),
\end{eqnarray}
and
\begin{eqnarray}
\label{zq3}
 {\cal Z}_{q_3} & =  & \int[d \Psi_3]{\rm exp}\{\frac{1}{2}\sum_{n,{\vec p}}  
 ~ {\bar \Psi}_3\frac{G_0^ {-1}}{T}\Psi_3 \} \nonumber \\
& = & {\rm Det} ^{1/2}(\beta G_0^{-1}),
\end{eqnarray}
where the determinantal operation ${\rm Det}$ is to be carried out over the Dirac, color, 
flavor and the momentum-frequency space.
In Eqs. (\ref{zq12}) and (\ref{zq3}), we have defined the quark propagator 
in the normal phase
\begin{equation}
{\rm G_0}^{-1} = 
    \left( 
          \begin{array}{cc}
            \left[ G_0^{+} \right]^{-1}  &  0 \\  
            0 &  \left[ G_0^{-} \right]^{-1} 
              \end{array}  
             \right),
\end{equation}
with 
\begin{eqnarray}
[G_0^{\pm}]^{-1}=
(p_0 \pm \mu) \gamma_0 -{\vec \gamma}\cdot {\vec p} -m
\end{eqnarray}
and $p_0=i \omega_n$,  and the quark propagator in the color-breaking phase
\begin{equation}
{\rm G}^{-1} = 
    \left( 
          \begin{array}{cc}
            \left[ G_0^{+} \right]^{-1}  &  \Delta^{-} \\  
            \Delta^{+}  &  \left[ G_0^{-} \right]^{-1} 
              \end{array}  
             \right). 
\end{equation}

The Nambu-Gorkov propagator ${\rm G}(p)$ is determined from solving
$1={\rm G}^{-1}\, {\rm G}$, resulting in
\begin{equation} 
\label{S}
{\rm G} = \left( \begin{array}{cc}
  G^{+} &   \Xi^{-}  \\
 \Xi^{+}  & G^{-}
           \end{array}  \right) \,\, ,
\end{equation}
with the components
\begin{eqnarray}
G^{\pm}  \equiv \left\{ \left[ G_0^{\pm} \right]^{-1} - \Sigma^{\pm} 
\right\}^{-1} \,\,\, , \,\,\,\, \Sigma^{\pm} \equiv \Delta^{\mp}
\, G_0^{\mp}\, \Delta^{\pm} \,\, , \nonumber
\end{eqnarray}
\begin{eqnarray}
\Xi^{\pm} \equiv -  G^{\mp} \, \Delta^{\pm} \, G_0^{\pm} 
= -  G_0^{\mp} \, \Delta^{\pm} \, G^{\pm}.
\end{eqnarray}
Here all components depend on the 4-momentum $p^\mu$.  

In the case of the mass term $m=0$, i.e., both the current quark mass 
$m_0$ and the chiral condensate $< \sigma >$ are zero, the Nambu-Gorkov quark 
propogator has a simple form which could be derived from
the energy projectors for massless particles \cite{rischke1}\cite{rischke2}, 
while if there is a small mass term, the quark propagator 
can be expanded perturbatively around $m=0$, but its form is very complicated 
\cite{rho2}. In our case the quark mass term can not be 
treated perturbatively, we have to find a general way to deal with the massive 
quark propagator. 

Fortunately, we can evaluate a simple form for the massive
quark propagator by using the energy projectors for massive particles.
The energy projectors onto states of positive and negative energy for free massive 
particles are defined as
\begin{eqnarray}
\Lambda_{\pm}({\vec p})=\frac{1}{2}(1\pm\frac{\gamma_0({\vec \gamma}
\cdot{\vec p}+m)}{E_p}),
\end{eqnarray}
where the quark energy $E_p=\sqrt{{\vec p}^2+m^2}$.
Under the transformation of $\gamma_0$ and $\gamma_5$, we can get another 
two energy projectors $\tilde \Lambda_{\pm}$,
\begin{eqnarray}
\label{tilde}
\tilde \Lambda_{\pm}({\vec p})=
\frac{1}{2}(1\pm\frac{\gamma_0({\vec \gamma}\cdot{\vec p}-m)}{E_p})\ ,
\end{eqnarray}
which satisfy 
\begin{eqnarray}
\label{gamma}
 \gamma_0 \Lambda_{\pm}({\vec p}) \gamma_0=\tilde \Lambda_{\mp}({\vec p}), ~ ~
 \gamma_5 \Lambda_{\pm}({\vec p}) \gamma_5=\tilde \Lambda_{\pm}({\vec p}).
\end{eqnarray}

The normal quark propagator elements can be re-written as 
\begin{eqnarray}
\label{mass0}
G_0^{\pm} =   \frac{\gamma_0\tilde \Lambda_{+}}{p_0+E_p^{\pm}} + 
\frac{\gamma_0\tilde\Lambda_{-}}{p_0-E_p^{\mp}},
\end{eqnarray}
with $E_p^\pm = E_p \pm \mu$. The propagator has four poles, i.e., 
\begin{eqnarray}
p_0=\pm E_p^{-},  ~~~p_0=\mp E_p^{+},
\end{eqnarray}
where the former two correspond to the excitation energies of particles and holes, and the latter
two are for antiparticles and antiholes, respectively.

The quark propagator including the diquark condensate's contribution can be evaluated as
\begin{eqnarray}
\label{gpmagain}
G^{\pm} = (\frac{p_0-E_p^{\pm}}{p_0^2-{E_{\Delta}^{\pm}}^2}\gamma_0\tilde\Lambda_{+}
+ \frac{p_0+E_p^{\mp}}{p_0^2-{E_{\Delta}^{\mp}}^2}\gamma_0\tilde\Lambda_{-})(\delta_{\alpha\beta}
-\delta_{\alpha 3}\delta_{\beta 3})\delta^{ij}  
\end{eqnarray}
and
\begin{eqnarray}
\label{cosiagain}
\Xi^{\pm}=(\frac{\Delta^{\pm}}{p_0^2-{E_{\Delta}^{\pm}}^2}\tilde\Lambda_{+}
+\frac{\Delta^{\pm}}{p_0^2-{E_{\Delta}^{\mp}}^2}\tilde\Lambda_{-}),
\end{eqnarray}
with ${E_{\Delta}^{\pm}}^2={E_p^{\pm}}^2+\Delta^2$. This propagator is very similar to
the massless propagator derived in \cite{rischke1}.

From the Nambu-Gorkov propagator, it is difficult to obtain the mass of the quark which participates in
the diquark condensate. The four poles of the Nambu-Gorkov propagator, i.e., 
\begin{eqnarray}
p_0=\pm E_{\Delta}^{-},  ~~~p_0=\mp E_{\Delta}^{+},
\end{eqnarray}
correspond to the excitation energies of quasi-particles (quasi-holes) and 
quasi-antiparticle (quasi-antiholes)
in the color breaking phase. These quasiparticles are superpositions of
particles and holes. 

\begin{figure}[ht]
\centerline{\epsfxsize=13cm\epsffile{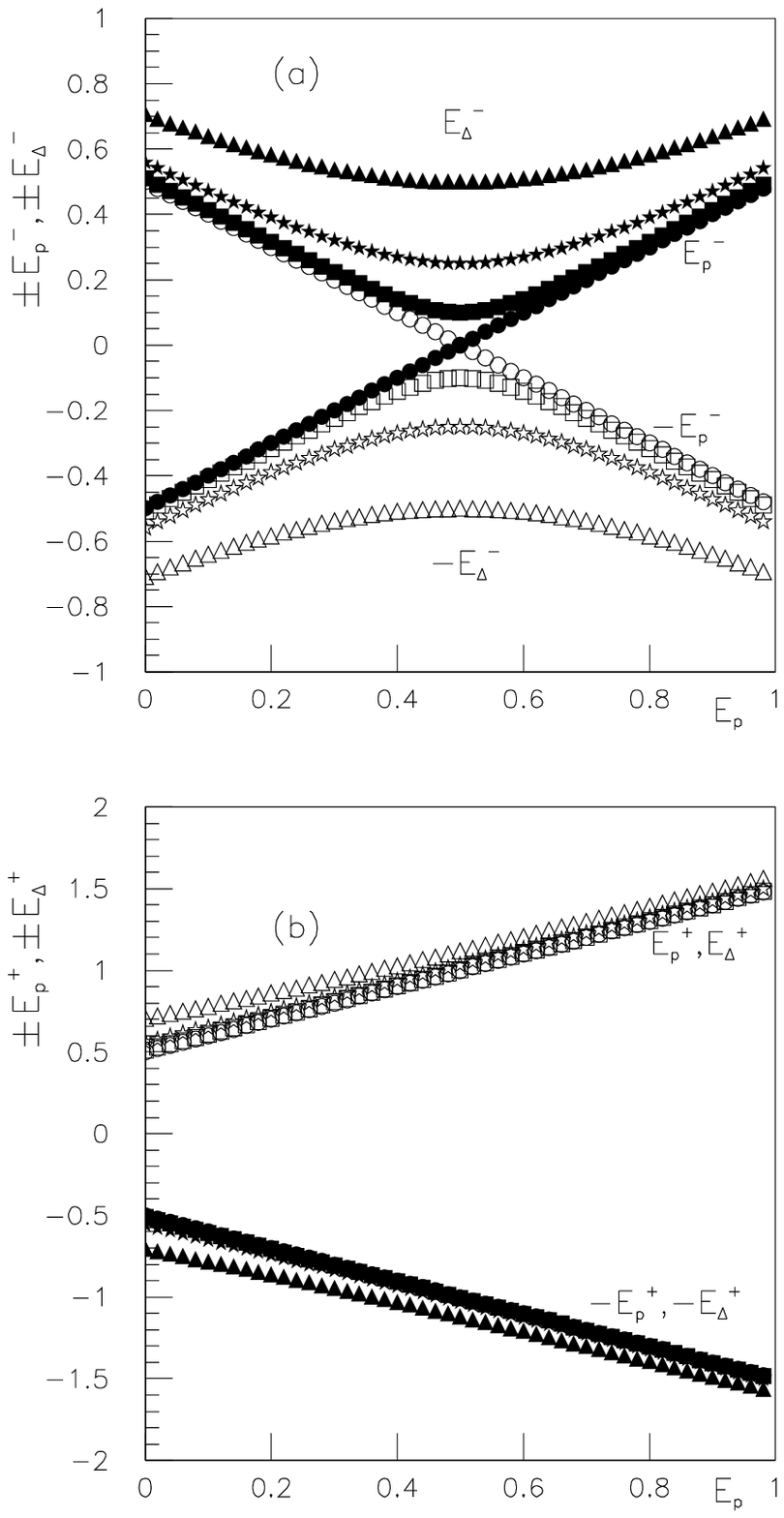}}
\caption{The excitation spectrum at $\mu=500 {\rm MeV} $ for (quasi-)particles
and (quasi-)holes in $(a)$ and for (quasi-)antiparticles and (quasi-)antiholes 
in $(b)$ as a function of 
$E_p$ with different values of $\Delta / \mu$, $\Delta / \mu=0$ (circles),
$0.2$ (squares), $0.5$ (stars) and $1$ (triangles). 
The black and white points are for 
the particles and holes, respectively.}
\label{excit_fig}
\end{figure}

We plot the excitation spectrum at $\mu=500 {\rm MeV} $ as a function of 
$E_p$ with different values of $\Delta / \mu$ in Fig. \ref{excit_fig}. 
$\Delta / \mu=0$ (the circles)
correspond to the excitation spectrum in the normal phase,
$\Delta/ \mu=0.2$ (the squares), $0.5$ (the stars) and $1$ 
(the triangles) correspond to the excitation spectrum in the color superconducting 
phase, the black points are for the particles and the white points
are for the holes.  $(a)$ is for the (quasi-)particles $ E_p^-( E_{\Delta}^-)$
and (quasi-)holes  $ -E_p^-( -E_{\Delta}^-)$, and $(b)$ is for the 
(quasi-)antiparticles $ -E_p^+( -E_{\Delta}^+)$ and (quasi-)antiholes
$ E_p^+( E_{\Delta}^+)$.

It can be easily seen from $(a)$ that the quasi-particles and  the quasi-holes 
mix particles and holes, which are called "Bogoliubons". 
In the normal phase, to excite a pair of particle and hole on the Fermi surface 
does not need energy, 
while in the superconducting phase, to excite a pair of quasi-particle and quasi-hole
at least needs the energy $2\Delta$ when $E_p=\mu$. With increasing $\Delta$, 
it is found that to excite a quasi-particle 
needs a larger energy, and the difference between the excitation energies at $E_p \neq \mu$
and that at $E_p = \mu$ becomes smaller.
 
From $(b)$, we see that to excite  an antiparticle is much more difficult, 
and the diquark condensate has little effects on the  excitation spectrum of
(quasi-)antiparticles and (quasi-)antiholes.

\subsection{The calculation of ${\rm ln} {\cal Z}_{q_3}$ }

For the third quark which does not participate in the diquark condensate, 
from Eq. (\ref{zq3}), we have 
\begin{eqnarray}
{\rm ln} {\cal Z}_{q_3} = \frac{1}{2} {\rm ln} \{  {\rm Det}(\beta [G_0]^{-1})
 = \frac{1}{2} {\rm ln}[ {\rm Det} (\beta [G_0^+]^{-1})  {\rm Det} (\beta [G_0^-]^{-1})].
\end{eqnarray}
 
Using the Dirac matrix, we first perform the determinant in the Dirac space,
\begin{eqnarray}
{\rm Det} ~ \beta ~[G_0^{+}]^{-1} & = & 
{\rm Det} ~ \beta~ [ (p_0+\mu) \gamma_0 - {\vec \gamma} \cdot {\vec p} -m]  \nonumber \\
   & = & {\rm Det} ~ \beta \left( 
          \begin{array}{cc}
            (p_0+\mu)-m  &  {\vec \sigma} \cdot {\vec p}  \\  
           - {\vec \sigma} \cdot {\vec p} &  -(p_0+ \mu)-m 
              \end{array}  
             \right), \nonumber \\
 & = & - \beta^2 ~ [ (p_0+\mu)^2- E_p^2 ],
\end{eqnarray}
and in the similar way, we get 
\begin{eqnarray}
{\rm Det} ~ \beta ~[G_0^{-}]^{-1} & = & - \beta^2 ~ [ (p_0-\mu)^2- E_p^2 ].
\end{eqnarray}
After performing the determinant in the Dirac space, we have
\begin{eqnarray}
{\rm Det} ~ \beta ~[G_0^{+}]^{-1} {\rm Det} ~ \beta ~[G_0^{-}]^{-1} = 
  \beta^2 [ p_0^2- (E_p + \mu)^2 ]~\beta^2  [ p_0^2- (E_p -\mu)^2].
\end{eqnarray}

Considering the determinant in the flavor, color, spin spaces and momentum-frequency 
space, we get the standard expression 
\begin{eqnarray}
{\rm ln} {\cal Z}_{q_3}  =  N_f
   \sum_n \sum_{\vec p} \{ {\rm ln} ( \beta^2 [ p_0^2- (E_p + \mu)^2 ] ) 
   +  {\rm ln} ( \beta^2  [ p_0^2- (E_p -\mu)^2] ) \},
\end{eqnarray}
remembering that the color space for the third quark is one-dimensional.  

\subsection{The calculation of ${\rm ln} {\cal Z}_{q_{1,2}}$ }

It is more complicated to evaluate the thermodynamic potential for the quarks 
participating in the diquark condensate. From Eq. (\ref{zq12}), we have 
\begin{eqnarray}
\label{detg}
{\rm ln} Z_{q_{1,2}}=\frac{1}{2} {\rm ln} {\rm Det} (\beta G^{-1}).
\end{eqnarray}

For a $2 \times 2$ matrix with elements $A,B,C$ and $D$, we have the identity
\begin{eqnarray}
\label{identity}
{\rm Det} \left(  \begin{array}{cc}
            A  &  B  \\  
           C & D
              \end{array}  
             \right) = {\rm Det} ( -CB + CAC^{-1}D ) ={\rm Det}(-BC+DC^{-1}AC).
\end{eqnarray}
To prove the above equation, we have used 
\begin{eqnarray}
\left(  \begin{array}{cc}
            A  &  B  \\  
           C & D
              \end{array}  
             \right) \equiv \left(  \begin{array}{cc}
            0  &  B  \\  
           C & 0
              \end{array}  
             \right)  \left(  \begin{array}{cc}
             1 &  C^{-1}D  \\  
           B^{-1}A & 1
              \end{array}  
             \right) \equiv 
            \left(  \begin{array}{cc}
           BC^{-1}  &  AB^{-1}  \\  
           DC^{-1} & CB^{-1}          
              \end{array}  \right)       
           \left(  \begin{array}{cc}
              0  &  C  \\  
                  B & 0
              \end{array}  
             \right) .  
\end{eqnarray}
Replacing $A,B,C$ and $D$ with the corresponding elements of $G^{-1}$,   we have 
\begin{eqnarray}
{\rm Det} (\beta {\rm G}^{-1}) 
& = &\beta^2 {\rm Det} D_1  =  \beta^2 {\rm Det} [- \Delta^{+} \Delta^{-} 
+ \Delta^{+}[G_0^{+}]^{-1} [\Delta^{+}]^{-1}[G_0^{-}]^{-1}] \nonumber \\
& = & \beta^2 {\rm Det} D_2 =  \beta^2 {\rm Det} [- \Delta^{+} \Delta^{-} 
+ [G_0^{-}]^{-1} [\Delta^{-}]^{-1}[G_0^{+}]^{-1}\Delta^{-}].
\end{eqnarray}
Using the energy projectors $\tilde\Lambda_{\pm}$, we can work out $D_1$ and $D_2$ as
\begin{eqnarray}
D_1 & = &  \Delta^2 +  \gamma_5 [\gamma_0 (p_0-E_p^{-}) \tilde\Lambda_{+}
+ \gamma_0 (p_0+E_p^{+})\tilde\Lambda_{-}] \gamma_5 
[\gamma_0 (p_0-E_p^{+})\tilde\Lambda_{+}
+ \gamma_0 (p_0+E_p^{-})\tilde\Lambda_{-}]   \nonumber \\
& = &  - [ (p_0^2-{E_p^{-}}^2-\Delta^2)\tilde\Lambda_{-}+
       (p_0^2-{E_p^{+}}^2-\Delta^2)\tilde\Lambda_{+}], \nonumber \\
D_2 & = & - [ (p_0^2-(E_p^{-})^2-\Delta^2)\tilde\Lambda_{+}+
       (p_0^2-(E_p^{+})^2-\Delta^2)\tilde\Lambda_{-}].
\end{eqnarray}
Using the properties of the energy projectors, we can get 
\begin{eqnarray}
D_1D_2=[ (p_0^2-(E_p^{-})^2-\Delta^2)] ~[(p_0^2-(E_p^{+})^2-\Delta^2)]
=[p_0^2-{E_{\Delta}^{-}}^2][p_0^2-{E_{\Delta}^{+}}^2].
\end{eqnarray}
With the above equations, Eq. (\ref{detg}) can be expressed as
\begin{eqnarray}
{\rm ln} {\cal Z}_{q_{1,2}} & = &  \frac{1}{2} {\rm ln} [ {\rm Det} \beta G^{-1}] 
 =  \frac{1}{4} {\rm Tr} {\rm ln}[ \beta^2 D_1 \beta^2 D_2] \nonumber \\
& = &  \frac{1}{4} \{ {\rm Tr}{\rm ln} [\beta^2 (p_0^2-{E_{\Delta}^{-}}^2) ] + 
                 {\rm Tr}{\rm ln}[\beta^2  (p_0^2-{E_{\Delta}^{+}}^2) ] \} \nonumber \\
& = & 2 N_f \sum_n \sum_p \{ {\rm ln} [\beta^2 (p_0^2-{E_{\Delta}^{-}}^2 )]+ 
                      {\rm ln} [\beta^2 (p_0^2-{E_{\Delta}^{+}}^2 ) ] \}.
\end{eqnarray}

\subsection{The thermodynamic potential}
 
The frequency summation of the free-energy
\begin{eqnarray}
\label{lnzf}
{\rm ln}{\cal Z}_f  =  \sum_n{\rm ln}[\beta^2(p_0^2-E_p^2)]
\end{eqnarray}
can always be obtained by performing the frequency summation of the 
propagator $1/(p_0^2-E_p^2)$.
Differentiate Eq. (\ref{lnzf}) with respect to $E_p$:
\begin{eqnarray}
\frac{\partial {\rm ln}{\cal Z}_f}{\partial E_p}  =  -2 E_p \sum_n \frac{1}{p_0^2-E_p^2}
=\beta [1-2 {\tilde f}(E_p)], 
\end{eqnarray}
where $\tilde f(x) = 1/(e^{\beta x} +1 )$ is the usual 
Fermi-Dirac distribution function.
Then integrating with respect to $E_p$, one can get the free-energy 
\begin{eqnarray}
{\rm ln}{\cal Z}_f=\beta [E_p +2T {\rm ln}(1+e^{-\beta E_p})].
\end{eqnarray}

With the help of the above expression, and replacing 
\begin{eqnarray}
\sum_p \rightarrow V \int\frac{d^3p}{(2\pi)^3},
\end{eqnarray}
we get the expressions 
\begin{eqnarray}
{\rm ln}{\cal Z}_{q_3}=  N_f \beta V \int \frac{d^3 p}{(2\pi)^3} [E_p^{+} 
+2T{\rm ln}(1+e^{-\beta E_p^{+}}) + E_p^{-} +2T {\rm ln}(1+ e^{-\beta E_p^{-}})],
\end{eqnarray}
\begin{eqnarray}
{\rm ln}{\cal Z}_{q_{1,2}}=  2 N_f \beta V \int \frac{d^3 p}{(2\pi)^3} [E_{\Delta}^{+} 
+2T{\rm ln}(1+e^{-\beta E_{\Delta}^{+}}) + E_{\Delta}^{-} 
+2T {\rm ln}(1+ e^{-\beta E_{\Delta}^{-}})].
\end{eqnarray}

Finally, we obtain the familiar expression of the thermodynamic potential  
\begin{eqnarray}
\label{potential}
\Omega & = & -T\frac{ {\rm ln} Z}{ V } =
  \frac{\sigma^2}{4G_S}+\frac{\Delta^2}{4G_D} 
-2N_f \int\frac{d^3 p}{(2\pi)^3} [ E_p + T{\rm ln}(1+e^{-\beta E_p^{+}}) 
+ T {\rm ln}(1+ e^{-\beta E_p^{-}})  \nonumber \\
& & + E_{\Delta}^{+} 
+2T{\rm ln}(1+e^{-\beta E_{\Delta}^{+}}) + E_{\Delta}^{-} 
+2T {\rm ln}(1+ e^{-\beta E_{\Delta}^{-}}) ].
\end{eqnarray}

\section{ Condensates and gap equations}

\subsection{Condensates}

With the Nambu-Gorkov quark propagator Eqs. (\ref{gpmagain}) and (\ref{cosiagain}), 
the diquark condensate is generally expressed as
\begin{eqnarray}
< {\bar q}^C \gamma_5 q>
= (iT\sum_n) \int\frac{d^3p}{(2\pi)^3}tr[\Xi^{-} \gamma_5].
\end{eqnarray}
From general consideration, there should be eight scalar diquark 
condensates \cite{rischke1}\cite{rischke3}. In the case of the NJL type model, 
the diquark condensates related to momentum vanish, and there
is only one independent $ 0^+$ diquark gap with Dirac structure 
$\Gamma=\gamma_5$ for massless quark, and there exists
another $0^+$ diquark condensate with Dirac structure
$\Gamma=\gamma_0\gamma_5$ at nonzero quark mass.
In our paper, we assume the contribution of the diquark condensate with 
$\Gamma=\gamma_0\gamma_5$ is small, and only consider the diquark condensate
with $\Gamma=\gamma_5$.  

Performing the Matsubara frequency summation and taking the limit $T \rightarrow 0$,
we get the diquark condensate at finite chemical potential
\begin{eqnarray}
\label{ddt0}
< {\bar q}^C \gamma_5 q> = -2 \Delta N_c N_f 
\int\frac{d^3}{(2\pi)^3}[\frac{1}{2E_{\Delta}^{-}}
+\frac{1}{2E_{\Delta}^{+}}].
\end{eqnarray}

For the third quark, its chiral condensate can be evaluated by using the
quark propagator in the normal phase,
\begin{eqnarray}
<{\bar q}_3 q^3>  =   -iT \sum_n \int\frac{d^3p}{(2\pi)^3}tr[G_0^{+}] ,
\end{eqnarray}
while for the quarks participating in the diquark condensate, the chiral condensate 
should be evaluated by using the quark propagator in the color breaking phase,
\begin{eqnarray}
<{\bar q}_{1} q^{1}>  = -iT \sum_n \int\frac{d^3p}{(2\pi)^3}tr[G^{+}].
\end{eqnarray}

We have no explicit mass expression for the first two 
quarks which participate in the diquark condensate, the influence of diquark condensate 
has been reflected in the quark propagator. The difference between the first two 
quarks which participate in the diquark condensate and the third quark 
which does not participate in the diquark condensate can be read from their chiral 
condensates and can be defined as 
\begin{eqnarray}
\label{delta}
\delta = <{\bar q}_1 q_1>^{1/3}-<{\bar q}_3 q_3>^{1/3},
\end{eqnarray}
where $\delta$ has the dimension of energy.
In the case of chiral limit, the quark mass $m$ decreases to zero in the color superconducting 
phase, and the influence of the diquark condensate on quarks in different colors vanishes.

After performing the Matsubara frequency summation and taking the limit $T \rightarrow 0$, 
we get the expressions of the condensates at $\mu \neq 0$,
\begin{eqnarray}
\label{condensate}
<{\bar q}_3 q^3>  & =  & 4 m N_f \int\frac{d^3p}{(2\pi)^3} \frac{1}{2 E_p}[\theta(\mu-E_p)-1], \nonumber \\
<{\bar q}_1 q^1> & = & 4 m N_f \int\frac{d^3p}{(2\pi)^3} \frac{1}{2 E_p}[n_p^{+}-n_p^{-}],
\end{eqnarray}
where
\begin{eqnarray}
n_p^{\pm}=\frac{1}{2}(1 \mp \frac{E_p^{\mp}}{E_{\Delta}^{\mp}})
\end{eqnarray}
are the occupation numbers for quasi-particles and quasi-antiparticles at $T=0$. 
Correspondingly, $1-n_p^{\pm}$ are the occupation numbers of quasi-holes and 
quasi-antiholes, respectively. 

\begin{figure}[ht]
\centerline{\epsfxsize=13cm\epsffile{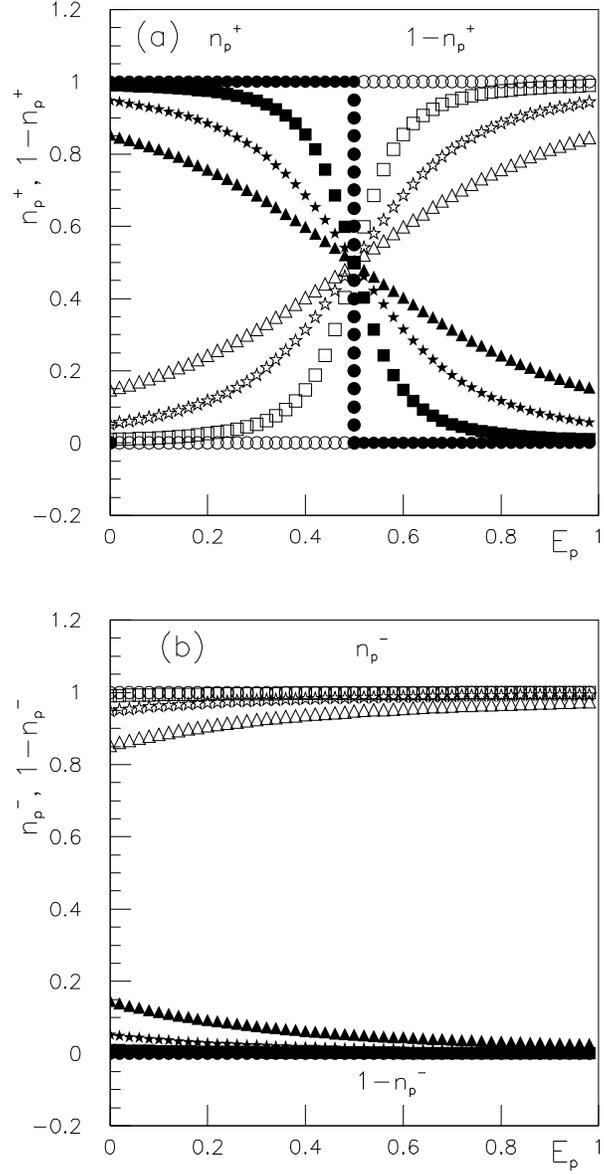}}
\caption{The occupation numbers for (quasi-)particles $n_p^{+}$  and (quasi-)holes
$1-n_p^{+}$ in $(a)$, and for (quasi-)antiparticles $n_p^{-}$ and (quasi-)antiholes 
$1-n_p^{-}$ in $(b)$ as a function of $E_p$ with
respect to $\Delta/\mu=0$ (circles), $0.2$ (squares), 
$0.5$ (stars) and $1$ (triangles). The black and white points correspond 
to particles and holes, respectively. }
\label{occn}
\end{figure}

We plot the occupation numbers for (quasi-)particles $n_p^{+}$  and (quasi-)holes
$1-n_p^{+}$ in Fig. \ref{occn} $(a)$, and the occupation numbers for (quasi-)antiparticles 
$n_p^{-}$ and (quasi-)antiholes $1-n_p^{-}$ in $(b)$ as a function of $E_p$ with
respect to $\Delta/\mu=0$ (circles), $0.2$ (squares), 
$0.5$ (stars) and $1$ (triangles); the black and white points correspond 
to particles and holes, respectively. 

It is seen that the Fermi surface is very sharp in the normal phase $\Delta/\mu=0$,
and becomes smooth when diquark condensate appears.  
The smearing is a consequence of the fact that the "Bogliubons" are superpositions 
of particle and hole states. The smearing of the Fermi surface 
induces the chiral symmetry restoring at a smaller 
chemical potential. The larger the diquark condensate is, the smoother the Fermi 
surface will be.
 
From Fig. \ref{occn}$(b)$, it can be seen that the occupation numbers 
for the (quasi-)antiparticles (antiholes)
in the normal phase or color breaking phase are not sensitive to the magnitude 
of the diquark condensate.  

\subsection{Gap equations}

The two gap equations $m$ and $\Delta$ can be derived by minimizing the 
thermodynamic potential Eq. (\ref{potential}) with respect to $m$ and $\Delta$,
\begin{eqnarray}
\frac{\partial \Omega}{\partial m} = \frac{\partial \Omega}{\partial \Delta} =0.
\end{eqnarray}
Taking into account the general expressions for the diquark condensate and
chiral condensates, one can get the relations between the
chiral gap $m$ and the chiral condensate  $ <{\bar q}q> $, i.e., 
\begin{eqnarray}
\label{mgap}
m & = & m_0 + \sigma, \nonumber \\
\sigma & = & -2 G_S <{\bar q}q>,
\end{eqnarray}
here the chiral condensate should perform summation in the color space
\begin{eqnarray}
<{\bar q}q> =  2 <{\bar q_1}q^1> + <{\bar q_3}q^3>;
\end{eqnarray}
and the relation between the diquark gap $\Delta$ and the diquark condensate 
$ <{\bar q}^C\gamma_5 q> $, i.e., 
\begin{eqnarray}
\label{dgap}
\Delta  = -2G_D<{\bar q}^C\gamma_5 q>,
\end{eqnarray}
substituting Eq. (\ref{ddt0}) into the above equation,
the gap equation for the diquark condensate Eq. (\ref{dgap}) 
in the limit of $T \rightarrow 0$ can be written as
\begin{eqnarray}
\label{dgap1}
1 = 4 N_c N_f G_D \int\frac{d^3p}{(2\pi)^3} [\frac{1}{2E_{\Delta}^{-}}
+\frac{1}{2E_{\Delta}^{+}}].
\end{eqnarray}

\section{Numerical Results}

In this section, through numerical calculations, we will
investigate the phase structure along the 
chemical potential direction, analyze the competition mechanism between the 
chiral condensate and diquark condensate, and discuss the influence of the color
breaking on the quarks in different colors.

Before the numerical calculations, we should fix the model parameters. 
The current quark mass $m_0=5. 5 {\rm MeV}$, the Fermion momentum 
cut-off $\Lambda_f=0.637 {\rm GeV}$, and the coupling constant in color
singlet channel $G_S=5.32 {\rm GeV}^{-2}$ are determined by fitting pion properties.
The corresponding constituent quark mass in the vacuum is taken to be
$m(\mu=0)=330 {\rm MeV}$.
The coupling constant in the color anti-triplet channel  $G_D$ can in principle be 
determined by fitting the nucleon properties. In \cite{dnjl2} $ G_D/G_S \simeq 2.26/3$
was chosen by fitting the scalar diquark mass of $ \simeq 600 {\rm MeV}$ to 
obtain a realistic baryon 
mass in the order of $\simeq 900 {\rm MeV}$. In our case, to investigate the influence
of diquark condensate on the chiral phase transition, we will set 
$G_D/G_S=0, 2/3, 1, 1.2, 1.5$, respectively. 
 
\subsection{Phase structure at zero temperature}

First, we investigate the phase structure along the chemical potential 
direction with respect to different magnitude of $G_D/G_S$. 
In the explicit chiral symmetry breaking case, we define the point at which
the chiral condensate has maximum change as the critical chemical potential
$\mu_{\chi}$ for the chiral phase transition, and
the point at which the diquark condensate starts to appear as the 
critical chemical potential $\mu_{\Delta}$ for the color 
superconductivity phase transition. 

\begin{figure}[ht]
\centerline{\epsfxsize=16cm\epsffile{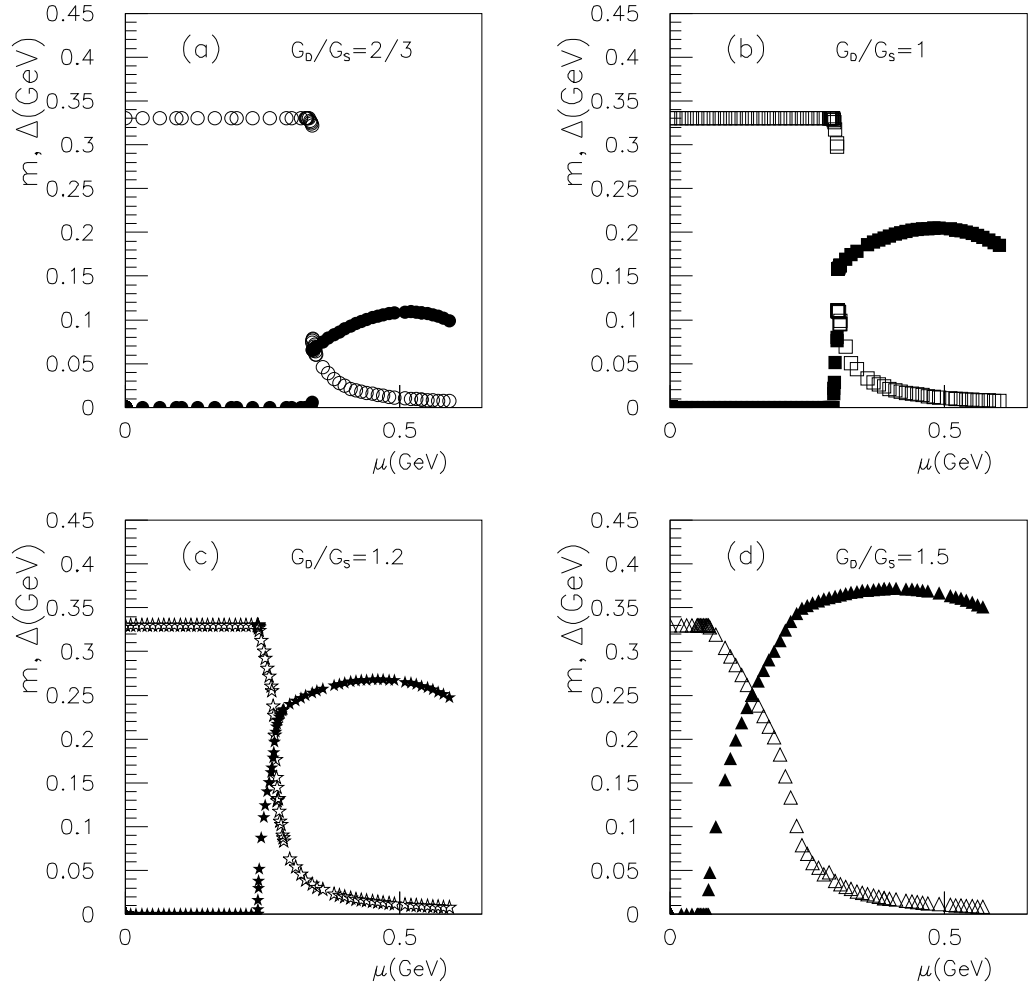}}
\caption{The two gaps $m$ (white points) and $\Delta$ (black points)
as functions of chemical potential $\mu$ for $G_D/G_S=0, 2/3, 1, 1.2, 1.5$,
respectively.}
\label{mqdd}
\end{figure}

\begin{figure}[ht]
\centerline{\epsfxsize=16cm\epsffile{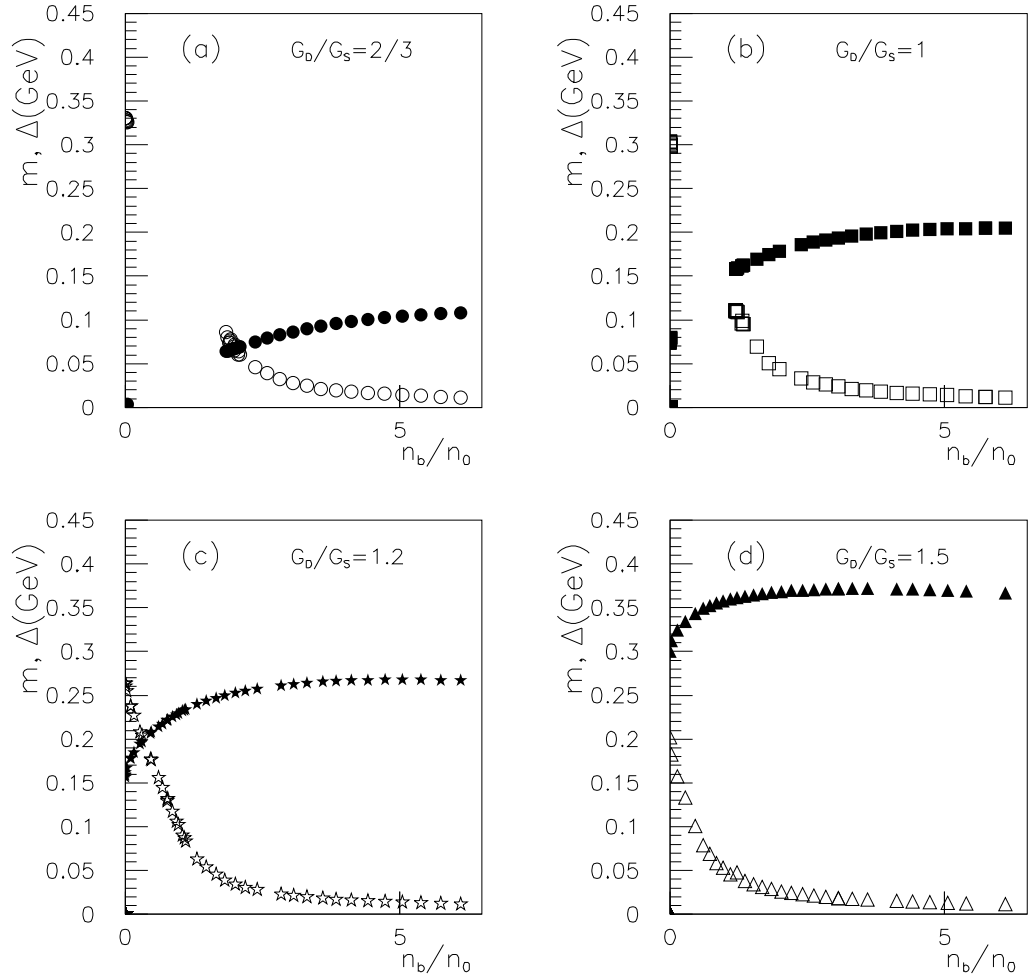}}
\caption{The two gaps $m$ (white points) and $\Delta$ (black points)
as functions of the scaled baryon density $n_b/n_0$ for
$G_D/G_S=0, 2/3, 1, 1.2, 1.5$, respectively.}
\label{nbmqdd}
\end{figure}

The two gaps $m$ (white points) and $\Delta$ (black points)
determined by Eqs. (\ref{mgap}) and (\ref{dgap}) are plotted in 
Fig. \ref{mqdd} as functions of $\mu$ with respect to different 
$G_D/G_S=2/3, 1, 1.2, 1.5$ in $(a), (b), (c)$ and $(d)$, respectively.
In Fig. \ref{nbmqdd}, they
are plotted as functions of the scaled baryon density $n_b/n_0$, where $n_0$
is the normal nuclear matter density.

In Fig. \ref{mqdd} $(a)$, i.e., in the case of $G_D/G_S=2/3$, 
we see that in the region where the constituent quark mass keeps 
its value in the vacuum $m(\mu=0)$, the diquark condensate keeps zero.
The chiral phase transition and the color superconductivity phase 
transition nearly occur at the same chemical potential 
$\mu_{\chi} \simeq \mu_{\Delta} = 340 {\rm MeV}$. The two phase
transitions are of first order. In the explicit chiral symmetry 
breaking case, there is a small chiral condensate in the 
color superconductivity phase $\mu>\mu_{\chi}$. This phenomena 
has been called the coexistence of chiral and diquark condensate 
in \cite{raja3}\cite{berges2}. In this coexistence region, 
the chiral condensate is small and can be described by the 
Andersom theorem \cite{kerbikov}.  
 
In $(b)$, with $G_D/G_S=1$, the diquark condensate starts to appear
at $\mu_{\Delta}=298 {\rm MeV}$, then chiral symmetry restores at 
$\mu_{\chi}=304.8 {\rm MeV}$. Both $\mu_{\Delta}$ 
and $\mu_{\chi}$ are smaller than those in the case of $G_D/G_S=2/3$.
In the region from $\mu_{\Delta}$ to $\mu_{\chi}$, both chiral and 
color symmetries are broken, which is called the mixed broken phase. 
The diquark gap increases continuously from zero to $82 {\rm MeV}$ 
in this mixed broken phase, and jumps up to $152 {\rm MeV}$ 
at the critical point $\mu_{\chi}$. The chiral
symmetry phase transition is still of the first order, and the jump of 
the diquark gap can be regarded as the influence of the first order 
chiral phase transition. 

In $(c)$ and $(d)$, we see that with increasing of $G_D/G_S$, 
the diquark condensate starts to appear at smaller $\mu_{\Delta}$, and 
chiral symmetry restores at smaller $\mu_{\chi}$, while the width of the region 
of mixed broken phase, $\mu_{\chi}-\mu_{\Delta}$, becomes larger. 
This phenomena has also been found in Ref.\cite{vander} in the random matrix model.
It is also seen that with increasing of $G_D/G_S$, the first order phase 
transition of chiral symmetry restoration becomes the second order one, 
where the critical point $\mu_{\chi}$ is still
defined by the maximum change of quark mass $m$. 

We summarize the phase structure along the chemical potential direction:
1) when $\mu< \mu_{\Delta}$, chiral symmetry is broken;
2) in the region from $ \mu_{\Delta}$ to $\mu_{\chi}$, both chiral and 
color symmetries are broken, and 3) when $\mu > \mu_{\chi}$, 
chiral symmetry restores partially and the phase 
is dominated by color superconductivity. The phase transition from chiral 
broken phase to the mixed broken phase is of second order, and the phase 
transition from the mixed broken phase 
to color superconductivity phase is of first order for 
small $G_D/G_S$, and of second order for large $G_D/G_S$.
The phase structure depends on the magnitude of $G_D/G_S$. If $G_D/G_S$ is very 
small, the diquark condensate will never appear; If $G_D/G_S<1$ but not too small,
there will be no mixed broken phase, the chiral phase transition 
and the color superconductivity phase transition
occur at the same critical point $\mu_{\chi}=\mu_{\Delta}$, 
and the two phase transitions are of first order like in \cite{raja3}.

\subsection{The Competition between the chiral and diquark condensate}

With increasing of $G_D/G_S$, the diquark condensate starts 
to appear at smaller $\mu_{\Delta}$ and chiral symmetry restores at smaller 
$\mu_{\chi}$, while the width of the mixed 
broken phase, $\mu_{\chi}-\mu_{\Delta}$, increases.

\begin{figure}[ht]
\centerline{\epsfxsize=14cm\epsffile{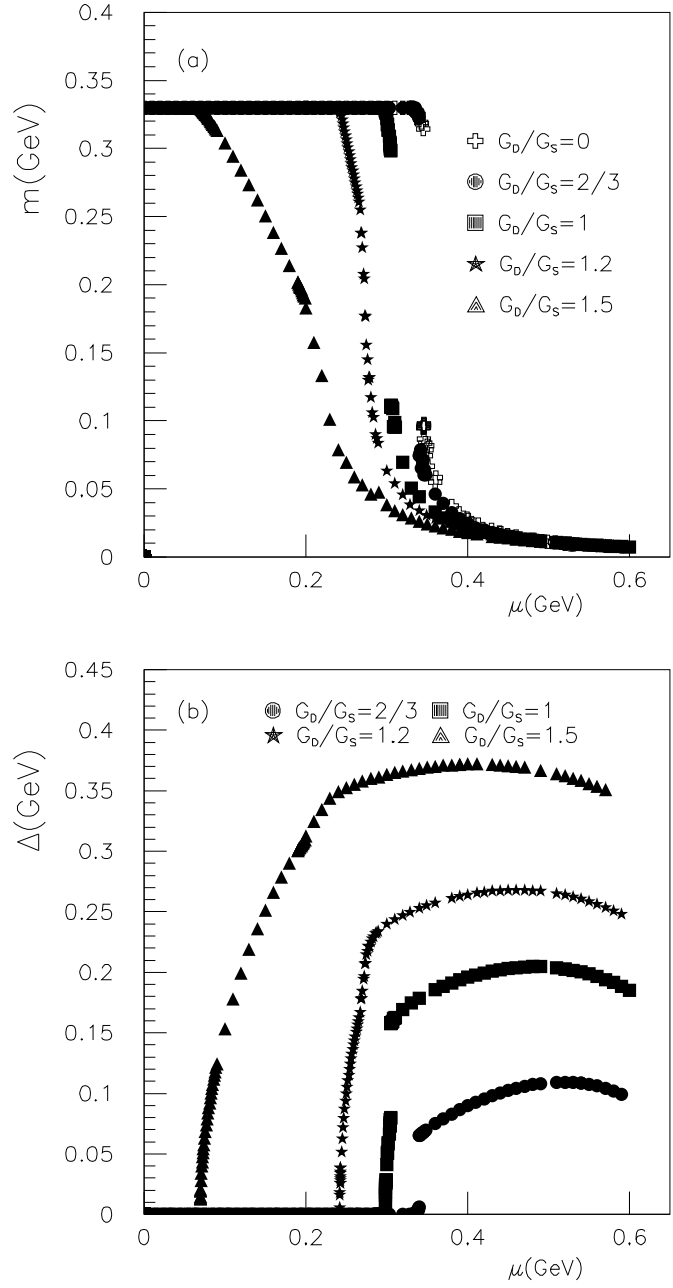}}
\caption{The gaps $m$ in $(a)$ and $\Delta$ in $(b)$ 
as a function of the chemical potential $\mu$ with respect to
$G_D/G_S=0, 2/3, 1, 1.2, 1.5$, respectively.}
\label{ddmq}
\end{figure}

In order to understand the competition mechanism and 
explicitly show how the diquark condensate influences the chiral phase
transition, we plot the constituent quark mass $m$ and the diquark
gap $\Delta$ as functions of $\mu$ for different values of 
$G_D/G_S$ in Fig. \ref{ddmq}. 

In the case of $G_D/G_S=0$, only chiral phase transition occurs,
the thermal system in the mean-field approximation is nearly a free 
Fermi gas made of constituent quarks. In the limit of $T=0$, there 
is a very sharp Fermi surface of 
the constituent quark.  When the chemical potential is larger than the 
constituent quark mass in the vacuum, the chiral symmetry restores, 
and the system of constituent quarks becomes a system of current quarks.
When a diquark gap $\Delta$ forms in the case of $G_D/G_S \neq 0$,  
it will smooth the sharp Fermi surface of the constituent quark.  
In other words, the diquark pair lowers the sharp Fermi surface, and 
induces a smaller critical chemical potential of chiral restoration.
 
In TABLE I, we list the chemical potentials $\mu_{\Delta}$, at which the 
diquark gap starts appearing, and $\mu_{\chi}$, at which 
the chiral symmetry restores, for different values of $G_D/G_S$.
$\mu_F^0=345.3 {\rm MeV}$ is the critical chemical potential
in the case of $G_D/G_S=0$. $\Delta_{\chi}$ is the value of diquark gap 
at $\mu_{\chi}$,  if there is a jump, it is the lower value. 

We see that for larger $G_D/G_S$, the diquark condensate
appears at a smaller chemical potential $\mu_{\Delta}$, and the chiral phase
transition occurs at a smaller critical chemical potential $\mu_{\chi}$,
the gap of the diquark condensate $\Delta_{\chi}$ at $\mu_{\chi}$ becomes larger, 
and the region of mixed broken phase becomes wider.

We assume the relation between $\Delta_{\chi}$ and  
$\mu_{\chi}$ as
\begin{eqnarray}
\mu_{\chi}=\mu_F^0 - x \Delta_{\chi}, 
\label{muchi}
\end{eqnarray}
and the relation between $\Delta_{\chi}$ and $\mu_{\Delta}$ as 
\begin{eqnarray}
\mu_{\Delta}=\mu_F^0 - y \Delta_{\chi}.
\label{mudelta}  
\end{eqnarray}

In TABEL I, we listed the values of $x=(\mu_F^0-\mu_{\chi})/\Delta_{\chi}$ and
$y=(\mu_F^0-\mu_{\Delta})/\Delta_{\chi}$ for different $G_D/G_S$.
It is found that $x$ is almost $G_D/G_S$ independent and equal to $1/2$.
As for $y$, it is larger than $1/2$
and increases with increasing of $G_D/G_S$. 
From the Eqs. (\ref{muchi}) and (\ref{mudelta}), we have the relation
\begin{eqnarray}
\mu_{\chi}-\mu_{\Delta}= (y-1/2) \Delta_{\chi}
\end{eqnarray}
for the mixed broken phase. 
With increasing of $G_D/G_S$, $y$ and $\Delta_{\chi}$ increase,
and then the width of the mixed broken phase, $\mu_{\chi}-\mu_{\Delta}$,
becomes larger. 

\begin{figure}[ht]
\centerline{\epsfxsize=14cm\epsffile{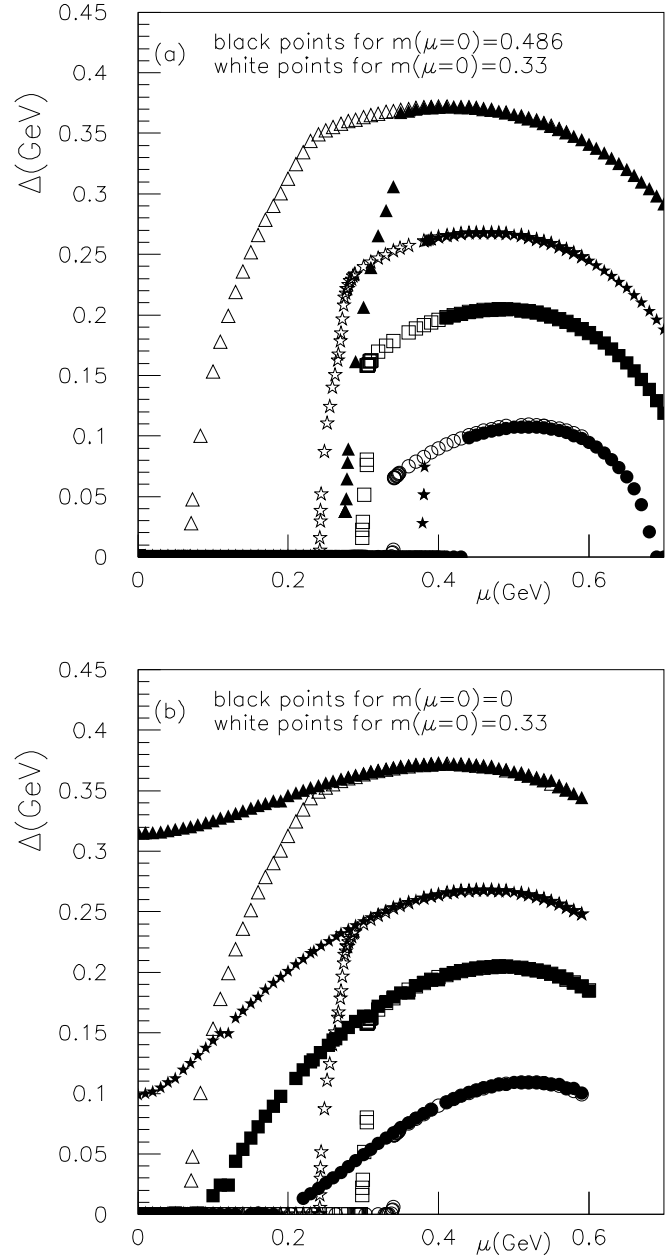}}
\caption{The influcence of chiral gap on the 
color superconductivity phase transition in the case of 
$m(\mu=0)= 486 {\rm MeV}$ in $(a)$ and $m(\mu=0)= 0$
in $(b)$. }
\label{gsingp}

\end{figure}

Now we turn to study how the chiral gap influences the color 
superconductivity phase transition. 
Firstly, we change the constituent quark mass in the vacuum from
$330 {\rm MeV}$ to $486 {\rm MeV}$.
To fit the pion properties, the coupling constant in the quark-antiquark channel
is correspondingly increased from $G_S$ to $1.2 G_S$.
We plot the diquark gap as a function of $\mu$ in Fig. \ref{gsingp}$a$ for
the two vacuum masses and for $G_D/G_S=2/3, 1, 1.2, 1.5$. 
We find that for the same $G_D/G_S$, the diquark gap starts to appear at
much larger chemical potential $\mu_{\Delta}$ when the vacuum mass
increases from $330 {\rm MeV}$ to $486 {\rm MeV}$.

Then we withdraw the quark mass, i.e., taking $m=0$ even in the vacuum.
We plot the diquark gap as a function of $\mu$ in Fig. \ref{gsingp} $b$ 
for $m(\mu=0)=0$ and $ 330 {\rm MeV}$ and for $G_D/G_S=2/3, 1, 1.2, 1.5$. 
We find that for any $G_D/G_S$, the diquark condensate starts 
to appear at a much smaller  
chemical potential $\mu_{\Delta}$ for $m(\mu=0)=0$ compared with
$m(\mu=0)=330 {\rm MeV}$.

From Fig. \ref{gsingp} $a$ and $b$, we can see that quark's vacuum mass
only changes the critical point of color superconductivity $\mu_{\Delta}$, 
the diquark gaps for different quark's vacuum mass 
coincide in the overlap region of color superconductivity phase,
where chiral symmetry restores partially.  

From the influcence of the diquark gap on the chiral phase transition
and the influence of the chiral gap on the color superconductivity phase 
transition, it is found that there does exist a strong competition between 
the two phases. The competition starts at $\mu_{\Delta}$ and ends at $\mu_{\chi}$.
We call the mixed broken phase as the competition region, which becomes wider 
with increasing of $G_D/G_S$. This competition region is the result of the  
diquark gap smoothing the sharp Fermi surface of the constituent quark.     
If the attractive interaction in the diquark channel is too small, 
there will be no diquark pairs, the system
will be in the chiral breaking phase before $\mu_{\chi}$, and in the chiral
symmetry restoration phase after $\mu_{\chi}$.
If the attractive interaction in the diquark channel is strong enough, 
diquark pairs can be formed and smooth the sharp Fermi surface,
and induce a smaller critical chemical potential of chiral
symmetry restoration.

\subsection{The influence of color breaking on quarks' properties}

Finally, we study how the diquark condensate influences the quark properties.

In the normal phase, the quarks in different colors are degenerate. 
However, in the color 
breaking phase, the first two quarks are involved in the diquark condensate, 
while the third one is not.

The quark mass $m$ appeared in the formulae of this paper is the mass for 
the third quark which does not participate in the diquark condensate. We have
seen from Fig. \ref{ddmq} $(a)$ that  the diquark condensate 
influences much the quark mass $m$ in the competition region 
$\mu_{\Delta} < \mu < \mu_{\chi}$. 
In the color breaking phase, i.e., when $\mu > \mu_{\Delta}$, 
the quark mass $m$ in different cases of $G_D$ decreases slowly with increasing 
$\mu$, and reaches the same value at about 
$\mu=500 {\rm MeV}$. 

\begin{figure}[ht]
\centerline{\epsfxsize=16cm\epsffile{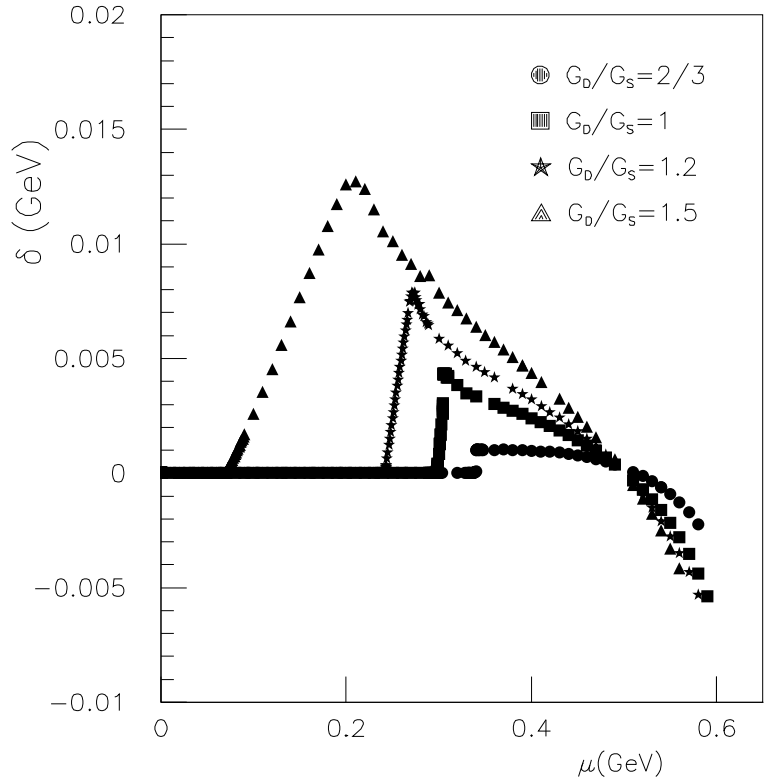}}
\caption{The difference of the 
chiral condensates for quarks in different colors $\delta$ as 
a function of the chemical potential $\mu$ with respect to
$G_D/G_S=0, 2/3, 1, 1.2, 1.5$, respectively.}
\label{qq13}
\end{figure}

The difference of the chiral condensates for quarks in different colors $\delta$ 
defined in Eq. (\ref{delta}) is shown in Fig. \ref{qq13} as a function of the 
chemical potential $\mu$ with respect to $G_D/G_S=2/3, 1, 1.2, 1.5$. 
It is found that in any case 
$\delta$ is zero before $\mu_{\Delta}$, then begins to increase 
at $\mu_{\Delta}$ and reaches
its maximum at $\mu_{\chi}$, and starts to decrease after $\mu > \mu_{\chi}$, and
approaches to zero at about $\mu=500 {\rm MeV}$,
when $\mu > 500 {\rm MeV}$, $\delta$ becomes negative.
With increasing $G_D$, $\delta(\mu_{\chi})$ increases from $1 {\rm MeV}$ 
for $G_D/G_S=2/3$ to $13 {\rm MeV}$ for $G_D/G_S=1.5$.
Comparing with the magnitude of the diquark condensate, $\delta$ 
is relatively small in the color superconductivity phase. 
 
\section{Conclusions}

In summary, in an extended NJL model and considering only the 
attractive interactions in the $0^{+}$ color singlet quark anti-quark 
channel and color anti-triplet diquark channel,
the Nambu-Gorkov form of the quark propogator has been 
evaluated with a dynamical quark mass. The Nambu-Gorkov massive propagator 
makes it possible to extend the Green-function approach to the moderate baryon 
density region, and the familiar expression of the thermodynamic potential 
has been re-evaluated by using the massive quark propagator. In this paper,
we have neglected another scalar diquark condensate with Dirac 
structure $\gamma_0\gamma_5$, which is assumed to be small.   

The phase structure along the chemical potential direction has been 
investigated.
The system is in the chiral breaking phasei before $\mu_{\Delta}$, 
in the color superconducting phase after $\mu_{\chi}$, 
and the two phases compete with each other in 
the mixed broken phase with width $\mu_{\chi}-\mu_{\Delta}$.
The width depends on the 
magnitude of $G_D/G_S$. If $G_D/G_S$ is small, the width
is zero, and the chiral phase transition
and the color superconductivity phase transition occur at the same
chemical potential. The width increases with increasing of $G_D/G_S$. 
 The competition mechanism has been analyzed 
by investigating the influence of the diquark condensate 
on the sharp Fermi surface. The diquark condensate smoothes 
the sharp Fermi surface, and induces the chiral phase transition 
occuring at a smaller chemical potential, and the diquark can be formed
more easily.
The phase transition is of second order from the chiral symmetry broken phase
to the mixed broken phase, and the phase transition from the mixed broken phase
to the partial chiral symmetry phase, i.e., the color superconductivity phase
is of first order for small $G_D/G_S$, and is of second order 
for large $G_D/G_S$. 

The influence of the diquark condensate on the properties of quarks in different colors has 
also been investigated. It is found that the difference of the chiral condensates
between quarks in different colors induced by the diquark condensate is very small.    

\section*{Acknowledgements}

One of the authors (M.H.) thanks valuable discussions with Dr. Qishu Yan.
This work was supported in part by China Postdoctoral Science Foundation, 
the NSFC under Grant Nos. 10105005, 10135030 and 19925519,  and the 
Major State Basic Research Development Program under Contract No. G2000077407.

\newpage

\newpage

\newpage

\newpage

\newpage

\newpage

\newpage

\newpage

\vspace*{5cm}
\begin{table}[ht]
\begin{center}
\begin{tabular}{|c|c|c|c|c|c|c|} 
$G_D/G_S$  &  $\mu_{\Delta} ({\rm MeV}) $ & $\mu_{\chi}  ({\rm MeV}) $  & 
$\mu_{\chi}-\mu_{\Delta} ({\rm MeV}) $ & $\Delta_{\chi}  ({\rm MeV})  $
& $ (\mu_F^0-\mu_{\chi})/\Delta_{\chi}  $  
& $ (\mu_F^0-\mu_{\Delta})/\Delta_{\chi}  $
\\ \hline
 1 & 298 & 304.8 & 6 & 82  &  0.49  &  0.57   \\  
\hline
 1.2 & 242 & 266 & 24 &   162   &  0.49   & 0.64   \\   
 \hline
 1.5 & 70 & 190 & 120 & 310  & 0.50  &  0.89  \\   
\end{tabular}
\vspace*{1cm}
\caption{ The $G_D$ dependence of chemical potentials $\mu_{\Delta}$ 
and $\mu_{\chi}$,  
$\mu_F^0=345. 3 {\rm MeV}$.}
\label{final_ex}
\end{center}
\end{table}

\end{document}